%% file: fg_chapter.tex
\documentclass[a4paper,openany,12pt]{book}
\include{defns_journals}

\usepackage[text={15.5cm,23cm},centering]{geometry}

\newcommand\savemathcal[1]{%
  \expandafter\newsavebox\csname mc#1content\endcsname%
  \expandafter\savebox\csname mc#1content\endcsname{$\mathcal{#1}$}%
  \expandafter\newcommand\csname mc#1\endcsname{%
    \expandafter\usebox\expandafter{\csname mc#1content\endcsname}}%
}
\newcommand\altmathcal[1]{\csname mc#1\endcsname}
\savemathcal{Z}
\savemathcal{B}

\usepackage{mathptmx}
\usepackage{graphicx}
\usepackage{chapterbib} 
\usepackage{subfigure}
\usepackage{color}
\usepackage{amsmath}
\usepackage{amssymb}
\usepackage[colorlinks=true, linkcolor=blue]{hyperref}
\usepackage[stable]{footmisc}

\usepackage[english]{babel}  

\begin{document}
\setcounter{chapter}{5}
\include{Chapman_Jelic/chapter}

\end{document}

%% file: Chapman_Jelic/chapter.tex
\chapter{Foregrounds and their mitigation\footnote{To appear as a book chapter in ``\textit{The Cosmic 21-cm Revolution: Charting the first billion years of our Universe}'', Ed. Andrei Mesinger (Bristol: IOP Publishing Ltd.) AAS-IOP ebooks  (http://www.iopscience.org/books/aas).}}
{\large \textbf{Emma Chapman} {(Imperial College London)}\\ 
{\large \textbf{Vibor Jeli\'c} (Ru{\dj}er Bo\v{s}kovi\'c Institute)}\\

\noindent \textbf{Abstract}\\
The low-frequency radio sky is dominated by the diffuse synchrotron emission of our Galaxy and extragalactic radio sources related to Active Galactic Nuclei and star-forming galaxies. This foreground emission  is much brighter than the cosmological 21~cm emission from the Cosmic Dawn and Epoch of Reionization. Studying the physical properties of the foregrounds is therefore of fundamental importance for their mitigation in the cosmological 21~cm experiments. This chapter gives a comprehensive overview of the foregrounds and our current state-of-the-art knowledge about their mitigation.

\section{What are the foregrounds?}
A detection of the redshifted 21~cm emission from the Cosmic Dawn (CD) and Epoch of Reionization (EoR) is a daunting task due to a number of challenges, which are different in nature and complexity. One of them is the extremely prominent foreground emission, which dominates the sky at low radio frequencies.  This emission intervenes like fog on an autumn morning and obscures our view towards the neutral hydrogen regions from the times of the first ``stars'' in the Universe. To clear the view and to make the detection possible, we need to study the foreground emission in great detail and acquire knowledge about its properties.

The foreground emission can be dived in two main categories: (i) Galactic foregrounds, mostly associated with the diffuse synchrotron and to some extent free-free emission from the Milky Way; and (ii) extragalactic foregrounds, associated with the radio emission from star-forming galaxies and Active Galactic Nuclei, and less relevant radio halos and relics. For an illustration of different foreground components see Fig.~\ref{fig:fgcube}. The former component dominates at angular scales larger than a degree and its contribution to the total foreground power is estimated to about 70\% at 150~MHz. The later component dominates at small angular scales and its contribution is estimated to about 30\%.  Both components are expected to be spectrally smooth due to the dominant synchrotron nature of their emission. 

In comparison to the cosmological 21~cm signal, the foreground emission is three to four orders of magnitudes brighter in total power. This amounts to two to three orders of magnitudes in fluctuations. Thus, the global redshifted 21~cm experiments, which use a single antenna for the measurement (e.g. EDGES), need to deal with an order of magnitude brighter foreground emission than the ones using interferometers (e.g. LOFAR, MWA and SKA). 

\begin{figure}[!t]
   \centering
    \includegraphics[width=0.95\textwidth]{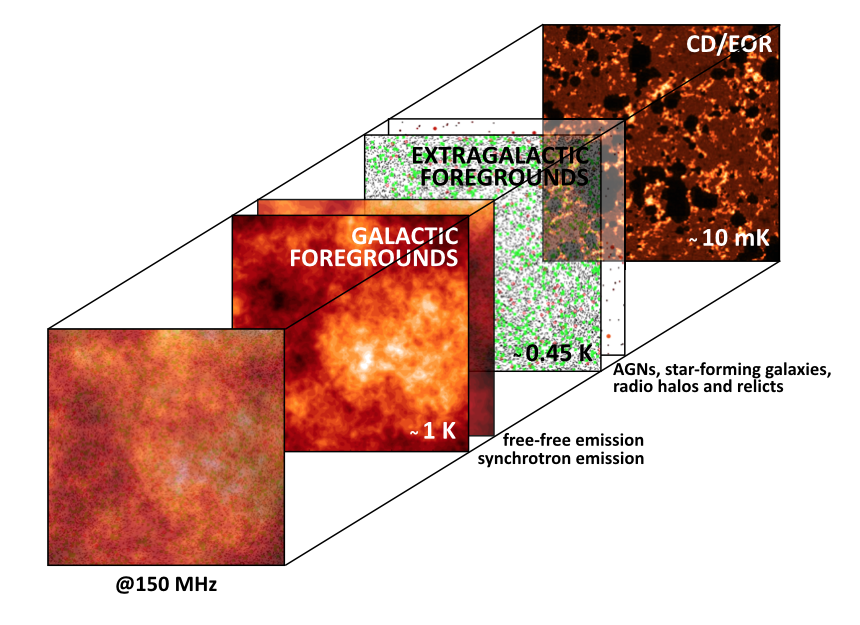}
    \caption{An illustration of different foreground components in the redshifted 21~cm experiments. The images are based on Jeli\'c simulations of the foregrounds \cite{jelic10, jelic08} and 21cmFAST simulations \cite{mesinger11}.}
    \label{fig:fgcube}
\end{figure}

The first overview of the foregrounds was outlined by \cite{shaver99}. Since then various authors have studied the foregrounds in the context of the cosmological 21~cm measurements \cite{bowman09, cooray04, deoliveiracosta08, dimatteo04, dimatteo02, jelic08, jelic10, liu12, oh03, petrovic11, spinelli18, wang06} (see also references in Sec.~\ref{sec:fgmit}).  
At the beginning these studies were mainly based on simulations shaped by extrapolated statistical properties of the foregrounds from the higher radio frequencies. The most comprehensive 
simulation of the foregrounds was carried by \cite{jelic08}.  This simulation has been used extensively in development of the robust foreground mitigation techniques for the LOFAR-EoR project \cite{Chapman2013MNRAS.429..165C, Chapman2012MNRAS.423.2518C, ghosh18,  Harker2010MNRAS.405.2492H, Harker2009MNRAS.397.1138H, Harker2009MNRAS.393.1449H, jelic08, mertens18} and more recently for the SKA CD/EoR project \cite{Chapman2015aska.confE...5C, Chapman2016MNRAS.458.2928C}. In addition to the dedicated foreground simulations, there are also more complex simulations of both Galactic and extragalactic emission, tailored for studies of the interstellar medium and magnetic fields in the Milky Way \cite{haverkorn19, sun09, waelkens09} or of different populations of the radio sources at low-radio frequencies \cite{bonaldi19, wilman10, wilman08}, that can be used as the foreground template in the cosmological 21~cm studies as well.

In parallel to the studies based on simulations, there were also a few dedicated observations taken with the WSRT \cite{bernardi09, bernardi10} and the GMRT \cite{pen09} radio telescopes to constrain the foregrounds at low-radio frequency. However, only once the new low-frequency instruments came online (e.g. EDGES, LOFAR, MWA and PAPER) our knowledge of the foregrounds started to grow extensively. In the following sections a more comprehensive overview of the foregrounds is given both in total intensity and polarization. 

\subsection{Galactic foregrounds in total intensity}
Galactic diffuse synchrotron emission is a dominant foreground component from a few tens of MHz to a few tens of GHz. It is non-thermal in its nature, produced mostly by the relativistic cosmic-ray electrons and to some extent positrons that spiral around the interstellar magnetic field lines and emit radiation.  Above a few tens of GHz free-free emission from diffuse ionized gas and thermal dust emission start to dominate over the synchrotron emission (see Fig.~\ref{fig:galcomp}). 

\begin{figure}[!t]
\centering
    \includegraphics[width=.95\textwidth]{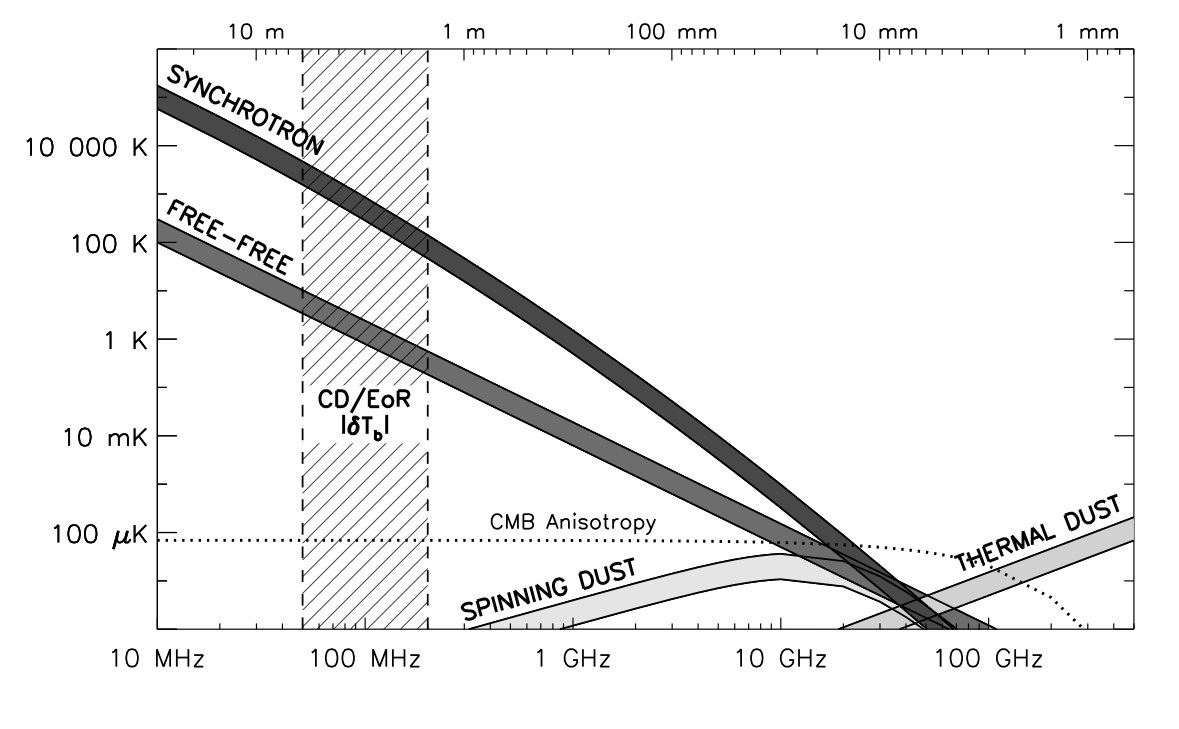}
    \caption{The main Galactic diffuse foreground components given as a function of frequency in the total intensity: (i) synchrotron emission from cosmic-ray electrons; (ii) free-free emission from diffuse ionized gas; and (iii) thermal dust emission. There is also a forth component associated with small rapidly spinning dust grains. Synchrotron emission dominates at frequencies below $\sim 10$ GHz, while thermal dust emission dominates at frequencies above $\sim 100$ GHz. Over the whole frequency range of the CD/EoR experiments, Galactic synchrotron emission is 3 -- 4 orders of magnitude stronger in total power (illustrated by the dark grey area) and 2 -- 3 orders of magnitude stronger in fluctuations than the cosmological 21~cm signal ($|\delta T_b|$). In the CMB experiments, on the contrary, there is a sweetspot around $~70$ GHz where the CMB anisotropies are relatively bright compared to the Galactic foreground emission.}
    \label{fig:galcomp}
\end{figure}

For a fairly complete theory of the synchrotron emission  please refer to e.g.~\cite{pacholczyk70,  rybicki86}, while here we outline the basics. The radiated synchrotron power emitted by a single electron is proportional to the square of the electron's relativistic kinetic energy, the magnetic energy density, and the pitch angle between the electron velocity and the magnetic field.  The angular distribution of the radiation is given by the Larmor dipole pattern in the electron's frame, but in the observer's frame is beamed sharply in the direction of motion. 

As the electron spirals around the magnetic field, it is in effect accelerating and emitting radiation over a range of frequencies. Its synchrotron spectrum has a logarithmic slope of 1/3 at low-frequencies, a broad peak near the critical frequency $\nu_c$, and sharp fall off at higher frequencies. The critical frequency is directly proportional to the square of the electron energy and the strength of the perpendicular component of the magnetic field. The longer the electron travels, the more energy it loses, the narrower spiral it makes, and the critical frequency is smaller. 

In the case of the Milky Way we need to take into consideration an ensemble of the cosmic-ray electrons, mainly originating from supernovae located close to the Galactic plane and then diffusing outwards. Given a typical magnetic field strength of a few $\mu$G, the cosmic-ray electrons with energies between 0.5 to 20 GeV account for the observed synchrotron radiation from tens of MHz to hundreds of GHz. Their energy distribution can be approximated with a power law with slope $\delta$:
\begin{equation}\label{eq:ncr}
n_{CR}(E)dE\propto E^{-\delta}dE,
\end{equation}
where $n_{CR}(E)dE$ is the number of cosmic-ray electrons per unit volume with energies between $E$ and $E+dE$. A distribution of their pitch angles is assumed further to be almost random and isotropic due to relatively long timescales (up to several millions of years) over which they lose their relativistic energies and due to repeatedly scattering that occurs in their environments. 

The observed synchrotron spectrum is then given by summing the emission spectra of individual electrons, which are smeared out in the observed spectrum by broad power law energy distribution of the comic-ray electrons. Thus, the synchrotron intensity at frequency $\nu$ depends only on $n_{CR}$ and $\delta$ from Eq.~\ref{eq:ncr} and on the strength of the magnetic field component perpendicular to the line-of-sight $B_\perp$:
\begin{equation}\label{eq:Isyn}
I_{\nu}\propto n_{CR}B_\perp^{(\delta+1)/2}\nu^{(1-\delta)/2}.
\end{equation}
The observed $I_{\nu}$ can be also described as a featureless power law in regards to the observed intensity $I_0$ at a reference frequency $\nu_0$:
\begin{equation}
I_{\nu}=I_0\left( \frac{\nu}{\nu_0} \right)^{-\alpha},
\end{equation}
where observed spectral index $\alpha$ is directly connected to the cosmic-ray  index $\delta$ as $\alpha=(\delta-1)/2$. Moreover,  the observed intensity is commonly expressed in terms of the brightness temperature $T_{b}(\nu)\sim\nu^{-\beta}$, using the Rayleigh-Jeans law which holds at radio frequencies. In this case the observed spectral index is $\beta=2+\alpha=2+(\delta-1)/2$. 

The comic-ray energy slope is estimated to  $-3.0<-\delta<-2.5$ at GeV energies \cite{lawson87, orlando13, strong11}. This corresponds to the synchrotron spectral index of $-1<-\alpha<-0.8$ or $-3<-\beta<-2.8$ observed at GHz frequencies \cite{platania98, reich88}.  At MHz frequencies the synchrotron spectrum is flatter \cite{guzman11, rogers08}. Typical values at mid and high Galactic latitudes are $-2.59 < -\beta < -2.54$ between 50 and 100 MHz \cite{mozdzen19} and $-2.62 < -\beta < -2.60$  between 90 and 190 MHz \cite{mozdzen17}, as measured by the EDGES instrument.  

A difference in the spectral index at MHz and GHz frequencies is due to ageing of the cosmic-ray energy spectrum. As the cosmic-ray electrons propagate trough the interstellar medium, they loose their energies by a number of energy loss mechanisms \cite{longair11} that involve interactions with matter, with magnetic fields and with radiation. This then depletes the population of relativistic electrons and changes their original energy (injection) spectra. For example, the energy loss trough synchrotron radiation is larger for cosmic-ray electrons with higher energies ($\sim E_{CR}^2$). The critical frequency is also proportional to $\sim E_{CR}^2$, so over time, the cosmic-ray spectra becomes steeper together with the synchrotron spectra at higher frequencies.  In a similar way, as the cosmic-ray electrons diffuse away from the Galactic plane, the ageing effect also makes a steepening of the synchrotron spectrum at higher Galactic latitudes \cite{strong07}.

Besides the spectral index variations across the sky, brightness temperature variations of the Galactic diffuse synchrotron emission reflect spatial fluctuations of the comic-ray electron density and magnetic field strength in the interstellar medium. Synchrotron emission is hence the brightest along the Galactic plane, which has the largest concentration of supernovae, a major source of the cosmic-ray particles, while the darkest parts are within the halo. This can be seen in Landecker all-sky map obtained at 150~MHz (see Fig.~\ref{fig:GDSE}, \cite{landecker70}), where typical high latitude brightness is between 150 K and 250 K. Given the low resolution of this map ($\sim5^\circ$), Haslam map at 408~MHz (see Fig.~\ref{fig:GDSE}, \cite{haslam82, haslam81, remazeilles15}) is more commonly used as a template for emission at low radio frequencies. 

\begin{figure}[!t]
\centering
    \includegraphics[width=0.475\textwidth]{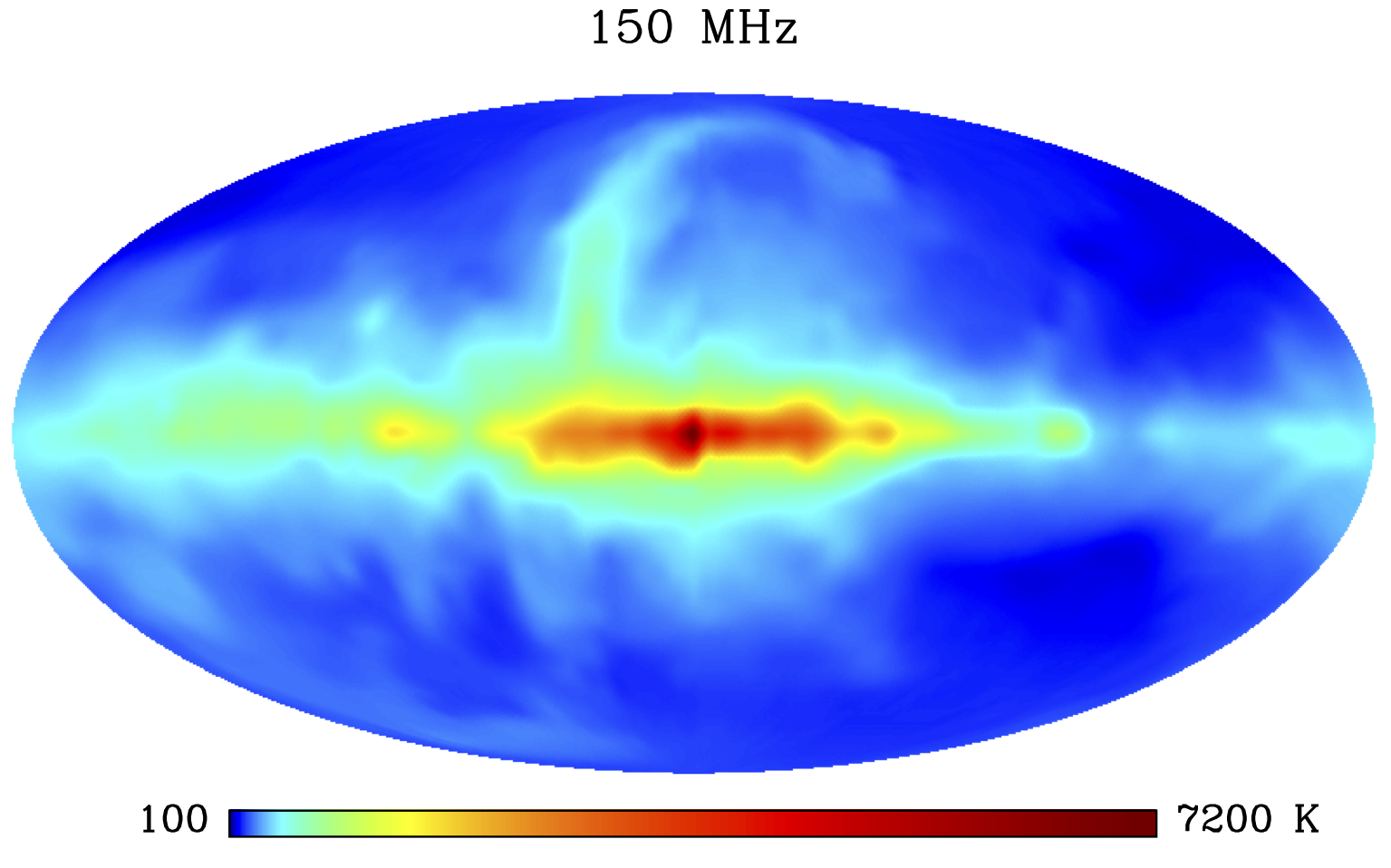}
    \includegraphics[width=0.475\textwidth]{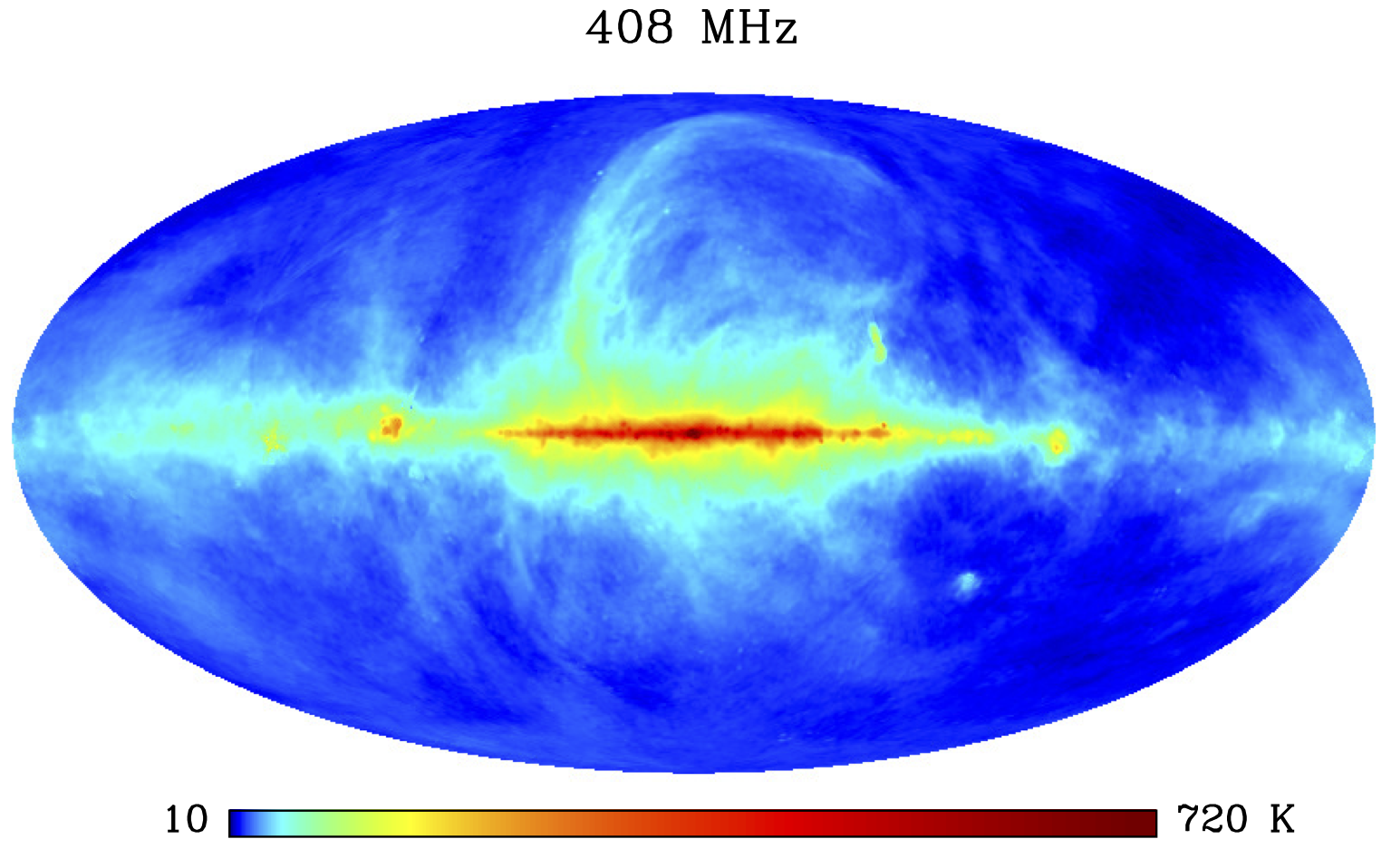}
    \caption{All sky maps of Galactic radio emission at 150 MHz \cite{landecker70} and 408 MHz \cite{haslam82, haslam81, remazeilles15}. \textit{This data is available on the Legacy Archive for Microwave Background Data Analysis (LAMBDA, https://lambda.gsfc.nasa.gov), a service of the Astrophysics Science Division at the NASA Goddard Space Flight Center.}}
    \label{fig:GDSE} 
\end{figure}

A number of recent dedicated observations additionally constrained Galactic synchrotron emission in selected areas at high Galactic latitudes. The WSRT observations at 150 MHz show an excess of power attributed to the diffuse synchrotron with an rms of 3 -- 5~K on scales greater than 30~arcmin (observations of the fields around 3C 196 and the North Celestial Pole,  \cite{bernardi10}). The LOFAR observation of the North Celestial Pole \cite{patil17} also clearly shows diffuse emission on scales larger than a degree, while slightly higher levels are found on scales greater than 54~arcmin in the MWA observations at 154~MHz of the fields near the South Galactic Pole \cite{lenc16}.

\subsection{Extragalactic foregrounds in total intensity}
Extragalactic radio sources are of composite nature.  They consist mainly of the active galactic nuclei (AGNs) or the star-forming galaxies (SFGs). 

Radio (synchrotron) emission in the AGNs, so called radio-loud AGNs, is related to the accretion of matter by a supermassive black hole at the centre of its host galaxy, typically an elliptical galaxy.  This produces narrow  jets in a direction perpendicular to the plane of the accretion. The jets can be as large as a few to ten times the size 
of the host galaxy and many of them have diffuse endings, so called radio lobes. Observed morphology of radio loud AGNs varies and can be classified in different ways. For example, we can classified them based on their radio luminosity and brightness of their components (nucleus, jets and lobes) \cite{fanaroff74}. In this case, the FR-I type galaxies  have lower radio powers with an edge darkened morphology, while the FR-II type galaxies have higher radio powers with an edge brightened morphology.

Radio emission in the SFGs is produced like in the Milky Way by synchrotron radiation from supernovae related relativistic electrons and by free-free emission from H{\sc ii} regions. Observed radio emission of these galaxies is usually also tightly connected, although still not well understood why, to the observed infrared luminosity measuring the star-formation rate (e.g.\cite{condon92, helou85, jarvis10}), hence the name SFGs. 

At low-radio frequencies different populations of radio galaxies are still poorly constrained, especially at the faint end of their distribution. There is a low-frequency extragalactic catalogue obtained with the MWA radio telescope in the south (GLEAM \cite{hurleywalker17}) and the ongoing LOFAR Two-metre Sky Survey (LoTTs \cite{shimwell17, shimwell19}) in the north. Until we get deeper with these surveys we need to rely on the data obtained at higher radio frequencies.

Normalised differential source counts for different populations of radio sources at 1.4 GHz is given in Fig.~\ref{fig:extragal}. Thanks to the recent very deep surveys (e.g. COSMOS, \cite{bondi08, novak18, smolcic17b, smolcic17a, smolcic17c}) the extragalactic radio sources are constrained well up to the flux densities of $500~{\rm \mu Jy}$. The population of the SFGs dominate at $\mu{\rm Jy}$ levels, while the population of the radio-loud AGNs dominates at flux densities $\geq1~{{\rm mJy}}$ (for a review see \cite{prandoniIAUS333} and references therein). There is also a third population of the sources detected below $\sim100~{\rm \mu Jy}$, commonly referred to as radio-quiet AGNs. These sources do not have large scale radio jets and lobes like radio-loud AGNs. They are probably SFGs hosting also an active nucleus that contributes to the radio emission \cite{ceraj18, delvecchio17}. 

\begin{figure}[!t]
   \centering
    \includegraphics[width=0.95\textwidth]{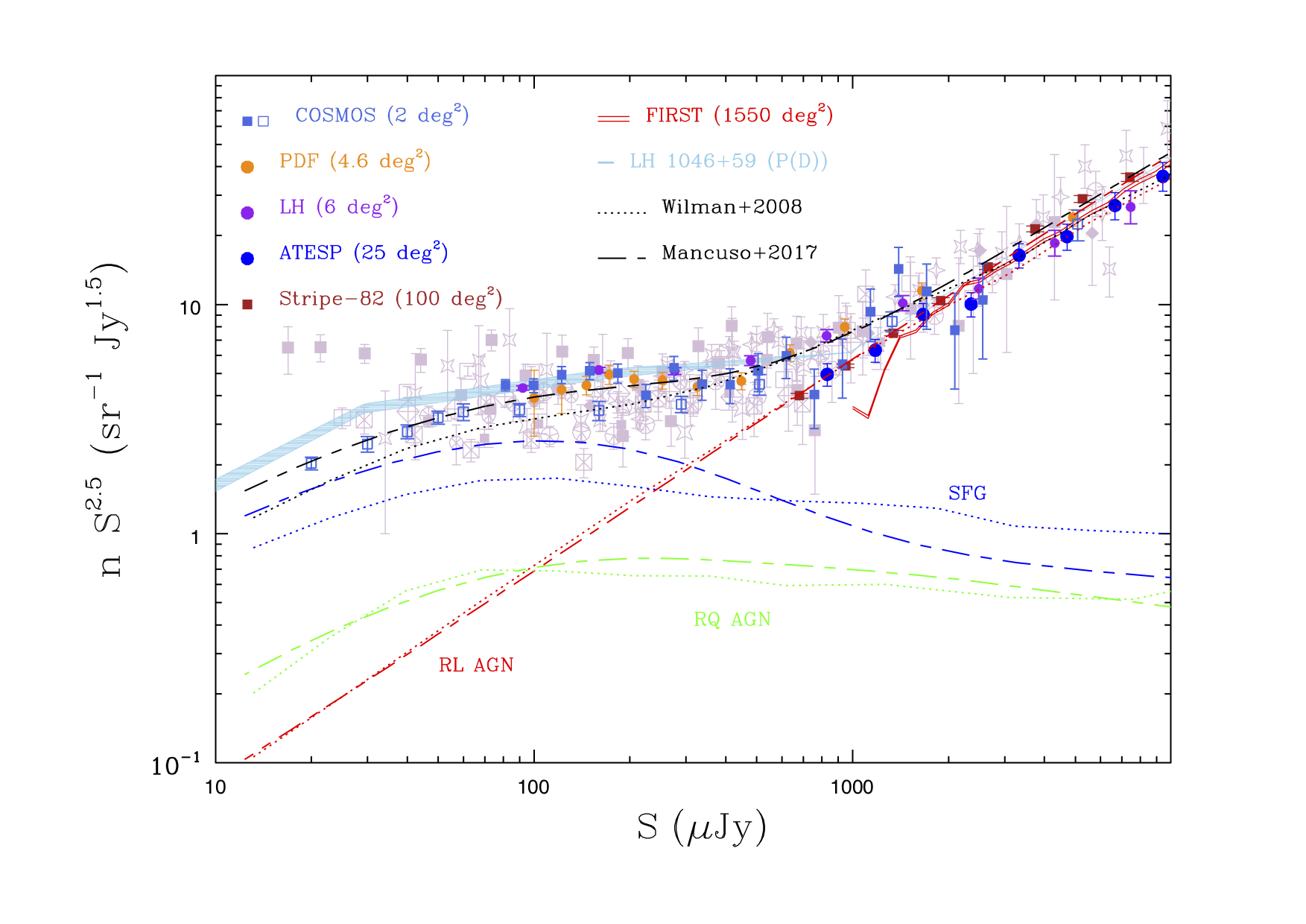}
    \caption{Normalized 1.4 GHz differential source counts. The dotted and dashed lines represent predicted counts from different model  \cite{mancuso17,wilman08,wilman10}. Different colours indicate different populations: radio-quite (RQ) AGN, radio-loud (RL) AGN and star-forming galaxies (SFG), while their sum is given in black. Coloured symbols show the counts from a number of large-scale surveys: COSMOS field \cite{bondi08, smolcic17b}; Phoenix Deep Field (PDF, \cite{hopkins03}); the Lockman Hole (LH, \cite{prandoni18}); the ATESP survey \cite{prandoni01}, the Stripe-82 region \cite{heywood16}; and the FIRST survey \cite{white97}. \textit{Reproduced from Prandoni 2018, Proceedings of the International Astronomical Union, IAUS 333:175--182}. }
    \label{fig:extragal}
\end{figure}

In addition to the radio source counts we also need to have a good knowledge of their distribution in the sky (clustering properties) and of their radio spectra. Neglecting source clustering may result in underestimating the angular foreground power which can potentially lead to a false detection of the cosmological 21 cm signal \cite{murray17,murray18}, while if the radio spectra is not smooth the foreground removal will be much more demanding.

The radio spectra of the radio galaxies can be described with the power-law function with a spectral index of  $\alpha\sim-0.7/-0.8$, due to the synchrotron nature of the emission.  Nevertheless, there are process that can change the shape of the spectra (free-free absorption, synchrotron self-absorption, spectral ageing, etc.) and make it complicated. Recent  LOFAR observations of the Bo\"otes field \cite{calistrorivera17} showed significant differences in the spectral curvature between SFG and AGN populations. The radio spectra of SFGs show a weak but statistically significant flattening, while the radio spectra of the AGNs is becoming steeper towards  the lower frequencies. Therefore, different power-law slopes should be assumed for AGNs and SFGs, when modelling the radio sky at frequencies relevant for the cosmological 21 experiments.

\subsection{Polarized foregrounds}\label{sec:polarfg}
Galactic synchrotron emission is partially linearly polarized. Its polarized intensity $PI_\nu$ depends on a cosmic-ray electron density $n_{CR}$, a slope of the cosmic-ray energy spectrum $\delta$, and a strength of the magnetic field component perpendicular to the line-of-sight $B_\perp$, in the same way as defined by Eq.~\ref{eq:Isyn} in total intensity. The only difference is the amount of emission, defined by the degree of polarization \cite{rybicki86}:
\begin{equation}
\Pi=\frac{\delta+1}{\delta+7/3}.
\end{equation}
For $\delta=2.2$, which is consistent with the observed synchrotron spectral index of $-\beta=-2.6$ at 150 MHz \cite{mozdzen17}, we get $\Pi=0.7$. At low radio frequencies  (100--200~MHz) about $70\%$ of Galactic synchrotron emission is intrinsically polarized, while in fact we observe only a few percent \cite{bernardi13, jelic14, jelic15, lenc17, lenc16, vaneck19, vaneck17}. To understand why we observe such a small percentage of polarized emission, we need to take a closer look at Faraday rotation and associated depolarization that occurs. 

As a linearly-polarized wave, with a wavelength $\lambda$, propagates through a magnetised plasma its polarization angle $\theta$ is Faraday rotated by:
\begin{equation}\label{eq:FR}
\frac{\Delta \theta}{\rm [rad]}= \frac{\lambda^2}{\rm [m^2]}\frac{\Phi}{\rm [rad~m^{-2}]}=  \frac{\lambda^2}{\rm [m^2]}\left(0.81\int \frac{n_e}{\rm [cm^{-3}]} \frac{B_{||}}{\rm [\mu G]} \frac{dl}{\rm [pc]}\right),
\end{equation}
where $\Phi$ is Faraday depth, $n_e$ is a density of the thermal electrons, $B_{||}$ is a strength of the magnetic field component parallel to the line-of-sight. The integral is taken over the entire path-length $l$, from the source to the observer.  The Faraday depth is positive when $B_{||}$ points towards the observer, while it is negative when $B_{||}$ points away.

In the Milky Way, where distributions of thermal and comic-ray electrons are perplexed throughout the entire volume, differential Faraday rotation will occur and will depolarize the observed synchrotron emission \cite{sokoloff98}.  As Faraday rotation is proportional to $\lambda^2$, depolarization at low radio frequencies will be significant. 
Nevertheless, small amounts of polarized emission that can still be observed carry valuable information about the physical properties of the intervening magnetised plasma.

\begin{figure}[!t]
\centering
    \includegraphics[width=0.475\textwidth]{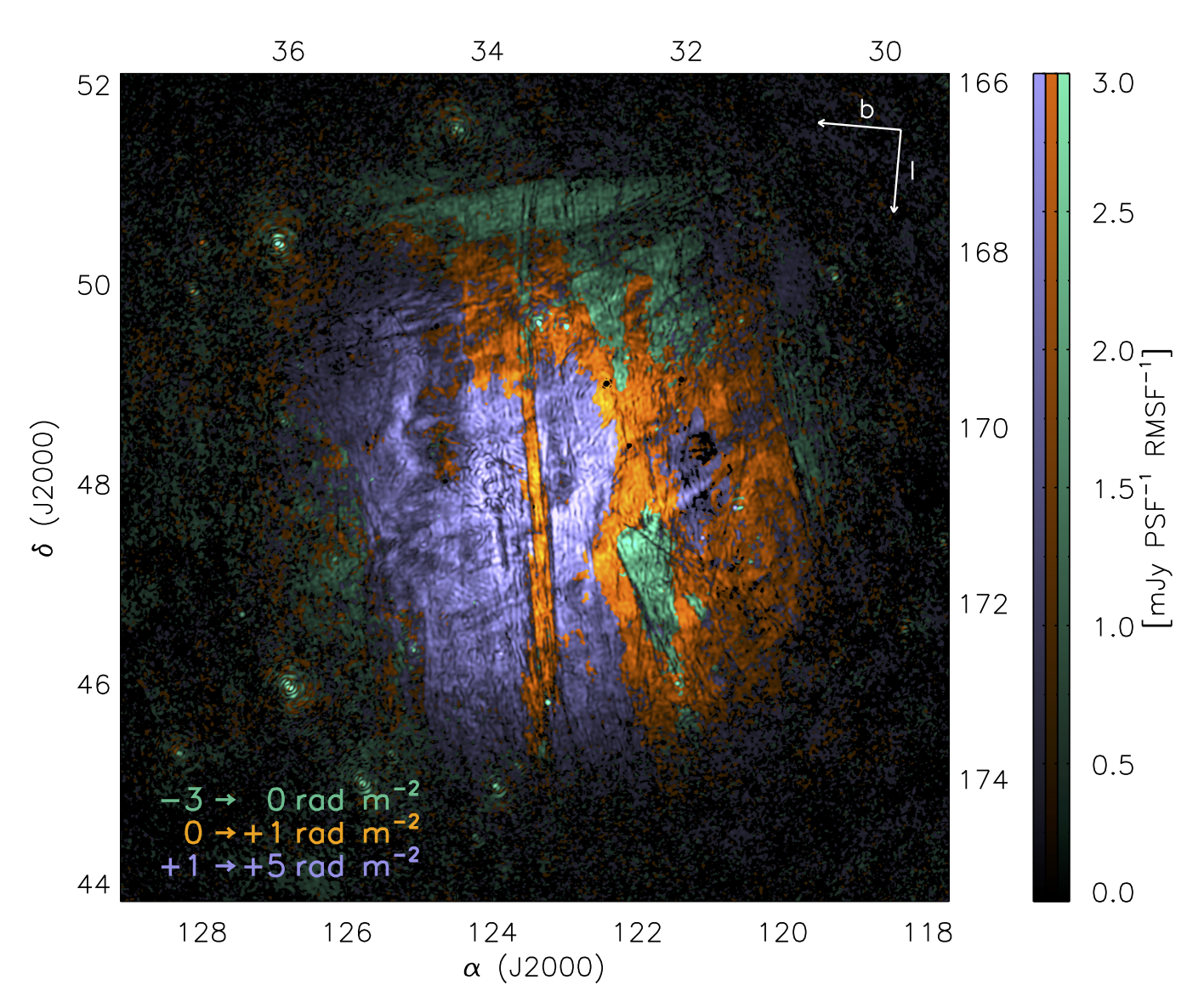}
    \includegraphics[width=0.475\textwidth]{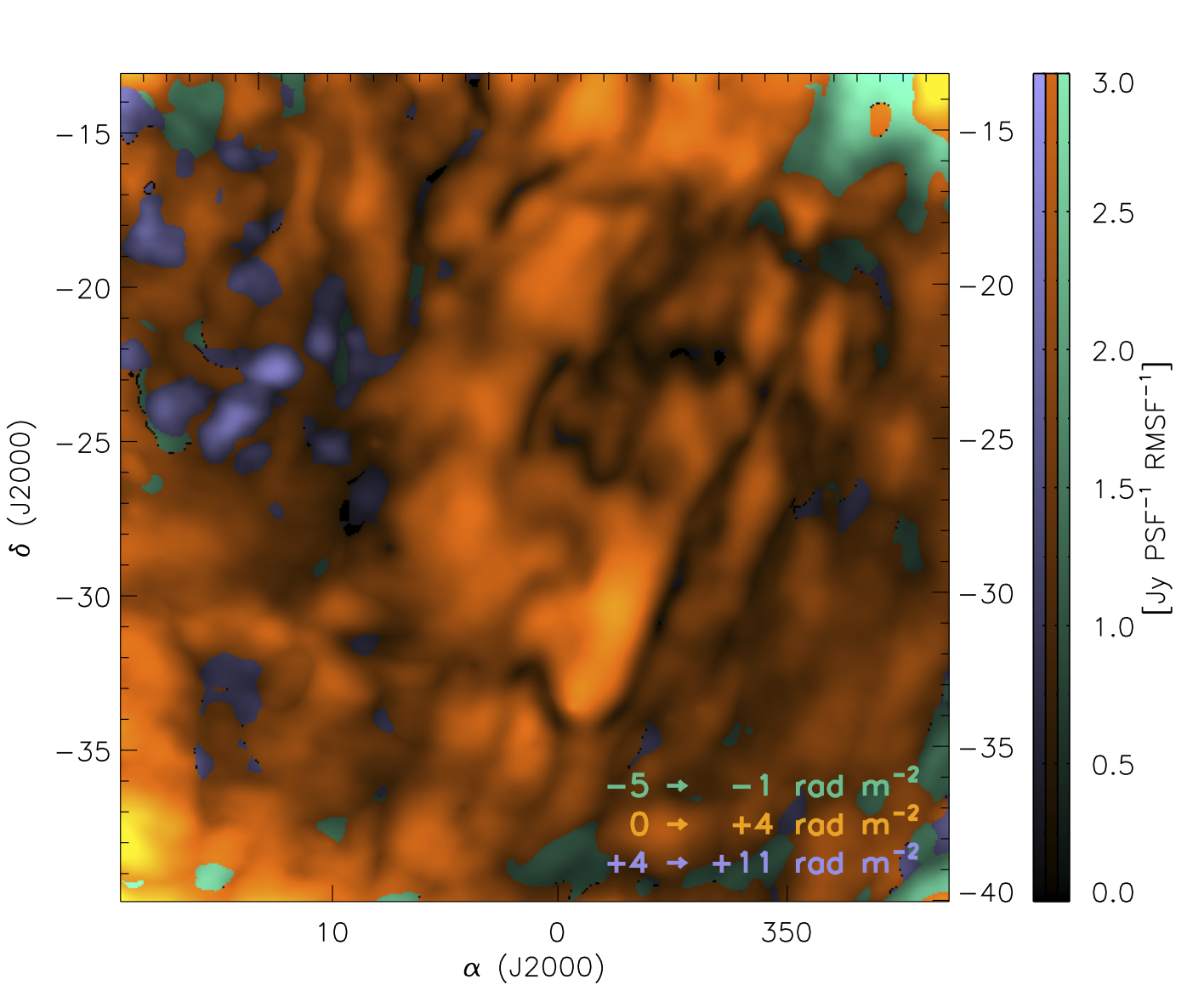}
    \caption{Polarized structures discovered at different Faraday depths with LOFAR (an image on the left - \textit{created using the data presented and discussed in Jeli\'c et al. 2015, A\&A, 583:A137, with permission of the authors}) and MWA (an image on the right -  \textit{created using the data presented and discussed in Lenc et al. 2016, ApJ, 830:38, with permission of the authors}) in two fields at high Galactic latitudes.}
    \label{fig:polar}
\end{figure}

First attempts to constrain diffuse polarized emission at 150 MHz were done using the GMRT \cite{pen09} and WSRT observations \cite{bernardi09, bernardi10}.  However, the full richness and complexity of polarized emission at low-radio frequencies was not revealed until LOFAR and MWA came online.  Observations with these instruments discovered astonishing morphology of polarized Galactic synchrotron emission of a few Kelvin in brightness (see Fig.~\ref{fig:polar} and \cite{bernardi13, iacobelli13b, jelic14, jelic15, lenc16, lenc17, vaneck19, vaneck17}). The discovered structures were unraveled by Rotation Measure (RM) synthesis \cite{brentjens05}. This is a technique in radio polarimetry that disentangles the observed wavelength-dependent polarization into a Faraday spectrum, i.e., the distribution of polarized emission as a function of Faraday depth. This allow us then to preform, so called, Faraday tomography, a study of the intervening magnetised plasma as a function of Faraday depth.

Given a wide frequency coverage and a high spectral resolution available in the low-frequency instruments Faraday tomography is preformed at an exquisite sensitivity and resolution in Faraday depth of $\sim1~{\rm rad~m^{-2}}$, an order of magnitude higher than at 350 MHz. This allow us to map small column densities of magnetised plasma that are, in most cases, not possible to detect at higher radio frequencies. Interestingly, most of the observed structures at low-radio frequencies appear at Faraday depths $\Phi\leq15~{\rm rad~m^{-2}}$ and they are not correlated with structures in total intensity. This result will be relevant in later discussion of the polarization leakage in the cosmological 21~cm experiments (see Sec.~\ref{sec:leakage}).

Extragalactic polarized sources are not a big concern for the cosmological 21~cm experiments due to their sparsity in the sky. In the MWA 32-element prototype survey of 2400 deg$^2$ of the southern sky at 189 MHz only one polarized source was found \cite{bernardi13}. In a preliminary data release of the LOFAR Two-meter Sky Survey of the HETDEX field, covering an area of 570 square degrees, 92 polarized radio sources where found \cite{vaneck18}. This gives a lower limit to the polarized source surface density at 150 MHz of only 1 source per 6.2 square degrees. Somewhat higher value, 1 source per 2 degrees, was found based on LOFAR observations of three 16 deg$^2$ fields \cite{jelic15, mulcahy14, vaneck18}.

\subsection{Radio Frequency Interference}
Terrestrially, radio frequency interference (RFI) from any human-made sources of radio transmission, such as wind turbines, leads to the necessary excision of frequency channels using a flagging technique (e.g. \cite{Offringa2012, Prasad2012ExA....33..157P}. The number of channels excised is significant, around 1$\%$ of channels of data for MWA and LOFAR \cite{Offringa2019MNRAS.484.2866O,Offringa2015PASA...32....8O}. Without careful mitigation in the calibration, imaging and diffuse foreground removal stages, RFI excision can result in an excess power that scales with the number of excised channels and does not integrate down with time, significantly dominating over the cosmological 21-cm signal by 1-2 magnitudes \cite{Offringa2019MNRAS.484.2866O}.
 
\section{Foreground Mitigation}\label{sec:fgmit}

The 21-cm signal emitted by high-redshift neutral hydrogen provides a window into the Epoch of Reionization (EoR), but it is a window that is obscured by layers of foregrounds. Extra-terrestrially, there exist a multitude of foregrounds which dominate all frequencies of observation and so more subtle methods than excision are required. This part of the chapter discusses the development and current use of Galactic and extragalactic foreground mitigation methods in Epoch of Reionization 21-cm experiments.

\subsection{Foreground Mitigation in the Data Analysis Pipeline}
\subsubsection{Bright Source Removal}
The first stage of foreground removal involves mitigating the effect of the very brightest sources on the sky: the point sources and extended sources. Bright source removal often comes under the umbrella of calibration as opposed to foreground mitigation however we will briefly summarize the process here. For example, the MWA real-time system (RTS) \cite{Mitchell2008ISTSP...2..707M} carries out sequential bright source `peeling' on the visibilities, tracking a few hundred of the brightest sources and comparing to a sky model constructed from existing catalogues and MWA observations \cite{Carroll2016MNRAS.461.4151C}. The gains are calibrated on the strongest source, before that source is peeled (subtracted) from the data, and the next strongest source is used to refine the calibration, and so on until it is deemed that enough bright sources have been removed, usually a few hundred to a thousand at most. The other MWA calibration pipeline, Fast Holographic Deconvolution (FHD) \cite{Sullivan2012ApJ...759...17S}, uses the MWA extragalactic catalogue GLEAM \cite{hurleywalker17} to calibrate gains, modelling all sources out to 1$\%$ beam level in the primary lobe, amounting to approximately 50000 sources \cite{Barry2019arXiv190102980B} and then removing a smaller population of them from the data. Similarly, LOFAR has built up a sky model over several years using the highest resolution LOFAR images and subtracts the sources in visibility space also \cite{Yata2015MNRAS.449.4506Y,Yata2013AA...550A.136Y}. As of 2017, the LOFAR EoR sky model contained around 20,800 unpolarized sources. 

\subsubsection{The EoR Window}
\label{sec:wedge}

It has previously been traditional when discussing diffuse foreground mitigation to assume that the previous stage of bright source subtraction has already been implemented perfectly. This is no longer seen to be a valid or safe assumption, as the chromaticity of the instrument, calibration errors and incorrect source subtraction lead to significant bias in the EoR signal for all current and planned experiments (e.g. \cite{ EW2017MNRAS.470.1849E,Procopio2017PASA...34...33P,Barry2016MNRAS.461.3135B,Patil2016MNRAS.463.4317P,Datta2010ApJ...724..526D,Liu2009MNRAS.394.1575L}, including redundant arrays \cite{Byrne2019ApJ...875...70B}). 

\begin{figure}
\begin{center}
    \includegraphics[width=0.5\textwidth]{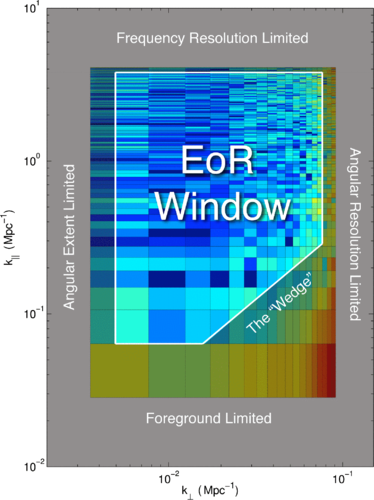}
\end{center}
    \caption{A schematic of the `EoR Window' in the cylindrical $k_{\parallel}$,$k_{\perp}$ Fourier plane, taken from Fig.1 of \cite{Dillon2014PhRvD..89b3002D}. In a perfect observation, with zero instrumental effects, the foregrounds would be entirely contained in the well defined horizontal band. In a realistic observation however, the chromaticity of the instrument results in a leakage of power up into the EoR window, into a region called the `wedge'. Aside from these contaminated areas there should be a relatively clean area called the EoR window. \textit{Reproduced from Dillon, Liu, Williams et al. (2014), PhysRevD, 89(2):023002}.} 
    \label{fig:window_liu}
\end{figure}

The spectral differences between the EoR signal and the bias introduced by the foregrounds and instrument lend themselves to a neat separation in $k_\bot-k_\parallel$ space, Fig. \ref{fig:window_liu}. In this formalism, spectrally-smooth foregrounds live in a well-defined area of $k$-space, at the smallest $k_\parallel$ scales, equivalent to the red stripe at the bottom of Fig. \ref{fig:window_liu}, excluding the wedge area. The assumption that the foregrounds would remain smooth and confined in a horizontal area at low $k_\parallel$ even after observation by a radio interferometer drove early foreground removal techniques such as those introduced in Section \ref{sec:poly} but is now known to be an incorrect assumption. The chromaticity of the instrument results in a `mode-mixing' where power is transferred from the angular to the frequency scales, throwing power upwards from the foreground area in the window into the larger $k_\parallel$ scales, with the effect increasing with larger $k_\bot$. This results in a wedge like structure, a structure that has been now extensively discussed and mathematically defined in the literature (e.g. \cite{Jensen2016MNRAS.456...66J,Dillon2014PhRvD..89b3002D,Liu2014PhRvD..90b3019L,Liu2014PhRvD..90b3018L,Hazelton2013ApJ...770..156H,Thyagarajan2013ApJ...776....6T,Pober2013ApJ...768L..36P,Morales2012ApJ...752..137M,Vedantham2012ApJ...745..176V,Trott2012ApJ...757..101T,Parsons2012ApJ...756..165P,Datta2010ApJ...724..526D}). Because the point sources reside on the largest $k_\bot$ scales they, or even their residuals when incorrectly calibrated, can overwhelm the EoR power in the frequency scales (e.g. \cite{Bowman2009ApJ...695..183B} and immediately preceding references).

Now we have defined the problem, namely the overpowering magnitude and potential leakage of foregrounds onto the EoR signal, we can consider how to achieve our aim of making accurate statistical conclusions on the nature of the EoR using the data within this window. To proceed, we can consider two philosophies. The first, \textbf{foreground subtraction}, aims to remove foreground contamination on all scales. The benefit of this is that there are more $k$ scales available for analysis. The drawback of foreground subtraction across all $k$-scales is that any failure in the method will potentially result in a foreground fitting bias across all scales of the window, providing another layer of contamination. One could instead avoid the foregrounds and therefore the need to remove them: \textbf{foreground avoidance}. This philosophy aims to then quantify the foregrounds and wedge such that any analysis occurs within a well-defined window free of contamination. The benefit of this is, as stated, the avoidance of foreground subtraction bias. The drawback is that any analysis is performed on a significantly reduced set of scales which can for example introduce its own bias into the spherically averaged power spectrum \cite{Jensen2016MNRAS.456...66J}. Additional to both philosophies, we can implement \textbf{foreground suppression}, which down-weights scales where the foregrounds or foreground removal residuals are dominant. We will now discuss these approaches in further detail in the context of current EoR experiments.

\subsection{Foreground Avoidance and Suppression}

The Murchison Widefield Array (MWA) has two separate pipelines which differ in their application of foreground mitigation techniques and calibration methods, while mostly employing foreground avoidance. The way in which MWA is optimized for making images allows the option to directly subtract known foregrounds but in this case the direct foreground subtraction is primarily applied to get access to a cleaner EoR window, not to get access to within the wedge, as is the motivation of foreground subtraction in LOFAR.

\begin{figure}
\begin{center}
    \includegraphics[width=0.9\textwidth]{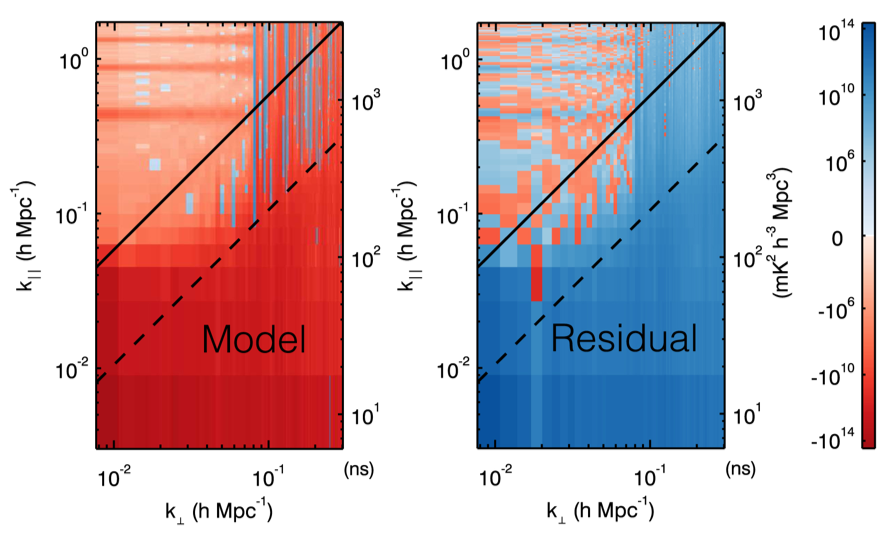}
\end{center}
\caption{Left: The difference between the MWA foreground model without diffuse foregrounds (i.e. just point sources) and with diffuse foregrounds. Adding diffuse foregrounds into the model produces leakage far up into the EoR window and instrumental contamination can be seen in the horizontal lines throughout the EoR window. Right: the difference between the power spectrum of the residuals when only the point sources have been subtracted as described above, and the power spectrum of the residuals where the diffuse foregrounds have also been subtracted. There is a clear reduction in foreground residuals all along the wedge and the EoR window is noise-like, suggesting a lack of foreground contamination there. There is a 70$\%$ reduction in residual power of the foregrounds using this method. \textit{Reproduced from Beardsley, Hazelton, Sullivan et al. (2016), ApJ, 833(1):102}.}
    \label{fig:beardsley_fg_sub}
\end{figure}

The FHD \cite{Sullivan2012ApJ...759...17S} and $\epsilon$psilon \cite{Barry2019arXiv190102980B} pipeline builds a sky model of point sources based on a golden set of data, including all sources above a floor limit within the primary beam of the instrument, and those beyond the primary beam if they are above 1$\%$ of the maximum primary beam level. This point source model is used in calibration in a similar way to LOFAR, and contains about 7000 sources as of 2016 \cite{Beardsley2016ApJ...833..102B}. In contrast to the RTS \cite{Mitchell2008ISTSP...2..707M} and CHIPS \cite{Trott2016ApJ...818..139T} pipeline, the FHD-$\epsilon$psilon pipeline also generates a diffuse foreground model by subtracting away the point source model from the observed data, and integrating over frequency to create a diffuse foreground model free of spectral information \cite{Beardsley2016ApJ...833..102B}. They then subtract both the point source model and the diffuse model from the data to minimise the leakage from the wedge into the EoR window. In Fig \ref{fig:beardsley_fg_sub} we see the effect of this foreground subtraction on the EoR window. The left image is the difference between the power spectrum of the MWA foreground model without diffuse foregrounds (i.e. just point sources) and with diffuse foregrounds. The plot shows that the diffuse foregrounds have power far up into the EoR window, due to non-uniform spectral sampling and the effect of windowing the data along frequency during the Fourier Transform. This figure if no other demonstrates the danger of assuming that the observed foreground signal is smooth and contained only at the smallest $k_\parallel$. Further instrumental complications can be seen in the horizontal lines throughout the EoR window, which is contamination due to the periodic frequency sampling function used by MWA \cite{Offringa2016MNRAS.458.1057O}. The right plot of Fig. \ref{fig:beardsley_fg_sub} shows the difference between the power spectrum of the residuals when only the point sources have been subtracted as described above, and the power spectrum of the residuals where the diffuse foregrounds have also been subtracted. There is a clear reduction in foreground residuals all along the wedge and the noise-like characteristic of the EoR window suggests a lack of foreground contamination there. \cite{Beardsley2016ApJ...833..102B} report a 70$\%$ reduction in residual power of the foregrounds using this method.

The black lines in Fig. \ref{fig:masks} show the area of the EoR window used in the FHD-$\epsilon$psilon pipeline, with the masks ensuring the avoidance of the horizontal contamination lines and the wedge.

\begin{figure}
\begin{center}
    \includegraphics[width=0.9\textwidth]{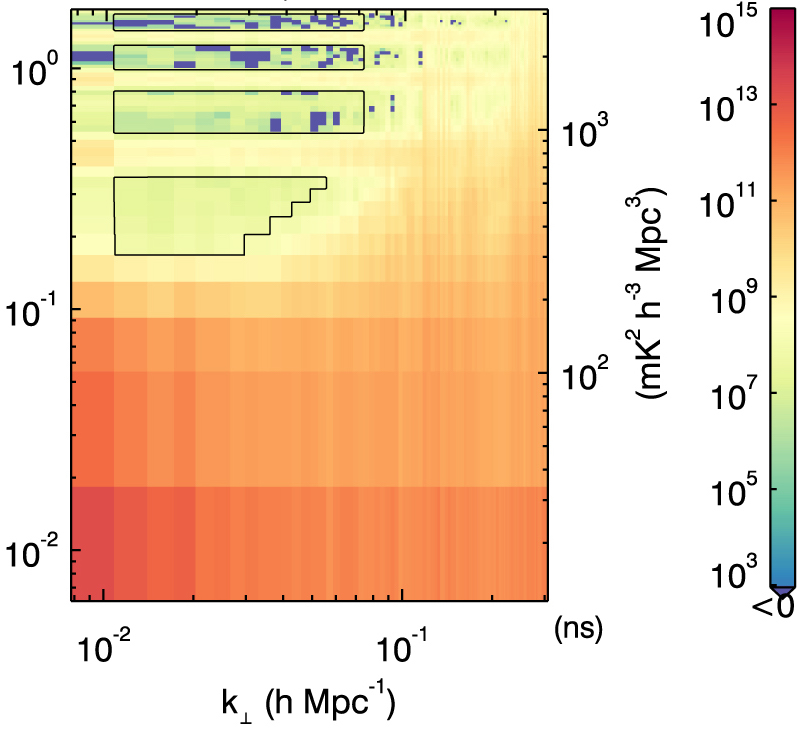}
\end{center}
    \caption{An example 2D cylindrical power spectrum of the first season MWA data after foreground mitigation. The data used for the upper limits can be seen bounded by black lines. The amount of data available for a power spectrum analysis has been severely reduced by the presence of foregrounds and instrumental contamination but the data within the bounded regions displays noise-like behaviour indicative of successful foreground mitigation. \textit{Reproduced from Beardsley, Hazelton, Sullivan et al. (2016), ApJ, 833(1):102}.}
    \label{fig:masks}
\end{figure}

The RTS-CHIPS pipeline subtracts significantly fewer sources, a few hundred to a thousand at most, and does so in visibility space. There is no diffuse foreground model in the subtraction stage and instead CHIPS down-weight modes with residual point source power. There is also the option of diffuse foreground weighting based on a simple foreground model where the covariances are known, though in practice this diffuse down-weighting is not currently utilised.

\subsubsection{Delay Space Filtering}
Delay space filtering is a method of foreground avoidance primarily adopted by the Donald C. Backer Precision Array for Probing the Epoch of Reionization (PAPER) \cite{Parsons2010AJ....139.1468P}. As with most foreground mitigation methods it requires the foregrounds to be reasonably smooth, even after instrumental effects. The wedge is the end-result of an instrument where the frequency-dependence of the instrument's sampling is dependent on the length of the baseline measuring the sky. Delay-space filtering exploits this relation by analyzing the data per baseline, circumventing the conspiracy of instrumental effects on the foregrounds and effectively isolating the foregrounds such that they are easily avoided. Fig. \ref{fig:baselines} demonstrates that for a given baseline measurement the visibility sampled changes with frequency, with a steeper change for longer baselines. This results in the mode-mixing seen in the 2D cylindrical power spectrum and the wedge structure, where we see power thrown up into the EoR window increasingly on the largest $k_\bot$ scales, which are the scales sampled by the longest baselines. Delay space filtering aims to mitigate the mode mixing by performing a Fourier transform along the visibility sampled by a given baseline (a solid line in Fig. \ref{fig:baselines}, and not along the frequency direction (vertical axis of Fig. \ref{fig:baselines}as is usual.

\begin{figure}
\begin{center}
    \includegraphics[width=0.5\textwidth]{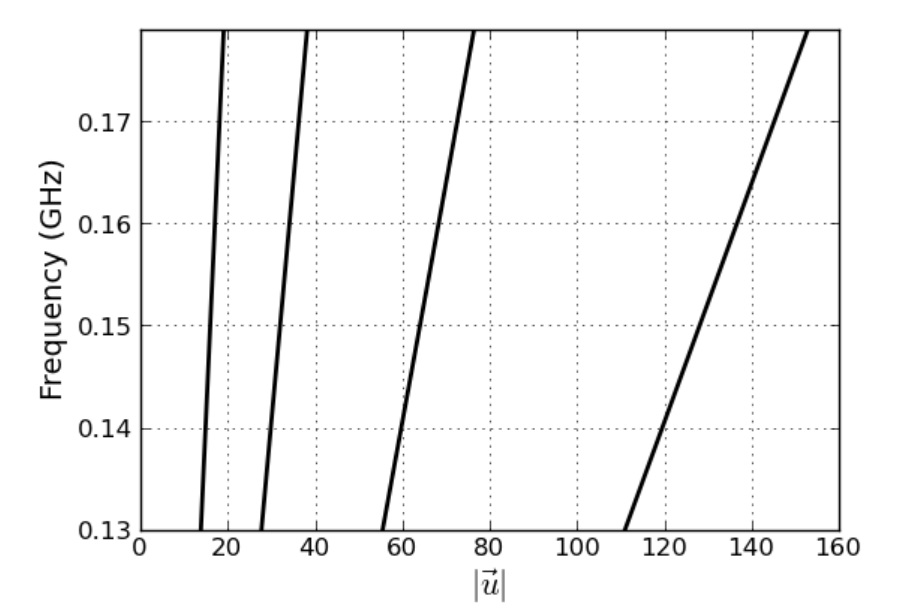}
\end{center}
    \caption{This figure demonstrates the frequency dependence of the wavemode sampled by baselines measuring 16, 32, 64, and 128 wavelengths at 150 MHz. For a given baseline measurement, the visibility sampled changes with frequency, with a steeper change for longer baselines. This results in the mode-mixing seen in the 2D cylindrical power spectrum and particularly "the wedge", where we see power thrown up into the EoR window increasingly on the smallest $k_\bot$ scales, which are the scales sampled by the longest baselines. \textit{Reproduced from Parsons, Pober, Aguirre et al. (2012), ApJ, 756(2):165}.}
    \label{fig:baselines}
\end{figure}

A delay transform takes a single time sample of a visibility from one baseline, for all observed frequencies (i.e. one of the solid lines on Fig. \ref{fig:baselines}, and Fourier transforms it to produce the delay spectrum \cite{Parsons2012ApJ...756..165P,Parsons2012ApJ...753...81P,Parsons2009AJ....138..219P}. The delay transform is:

\begin{equation}
    \widetilde{V}_b(\tau) = \int dl \; dm \; d\nu \; A(l,m,\nu)I(l,m,\nu)e^{-2\pi i\nu(\tau_g-\tau))}
\end{equation}

\noindent where $l,m$ have their usual definition relating to angular coordinates on the sky (e.g. \cite{Thompson2001isra.book.....T}). $\tau$ is the time-delay between the signal reaching both antennas and the geometric group delay associated with the projection of baseline $\overrightarrow{b} \equiv (b_x,b_y,b_z) $ in the direction $ \hat{s} \equiv (l,m,\sqrt{1-l^2-m^2})$ is:

\begin{equation}
    \tau_g \equiv \frac{\overrightarrow{b} \cdot \hat{s}}{c} 
\end{equation}

For comparison, the usual equation where the Fourier transform is simply applied along the frequency axis is:

\begin{equation}
    \widetilde{V}(u,v,\eta) = \int dl \; dm \; d\nu \; A(l,m,\nu)I(l,m,\nu)e^{-2\pi i(ul+vm+\eta\nu)}
\end{equation}

\noindent where $\eta$ is the Fourier transform of $\nu$.

The delay transform transforms flat spectra sky emission into delta functions. Because the sky emission is not perfectly smooth, and the instrument adds in its own unsmoothing effects, this delta function is effectively convolved with a kernel, which broadens the delta function in delay space. For the smoother foregrounds, that kernel will be narrow, and confined within the ``horizon limits", the geometric limit in delay space beyond which no flat spectra emission can enter the telescope. Spectrally unsmooth sky emission can enter beyond these horizon limits and emission such as the cosmological signal finds itself with a wide convolving kernel, spreading power well beyond the horizon limit where the foregrounds are theoretically confined. In Fig. \ref{fig:horizon} we see the delay transform at 150 MHz for several spectrally smooth sources and how they remain confined within the horizon limits of the baseline (here 32 metres). In contrast, the delay spectrum of spectrally unsmooth emission, such as the cosmological 21-cm signal, finds itself smeared to high delays. Full mathematical detail can be found in \cite{Parsons2012ApJ...756..165P,Parsons2012ApJ...753...81P} and \cite{Parsons2009AJ....138..219P}.

\begin{figure}
\begin{center}
    \includegraphics[width=0.5\textwidth]{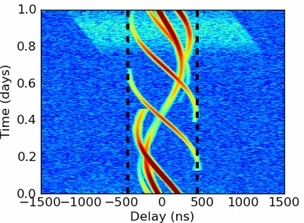}
\end{center}
    \caption{The delay spectra of several smooth-spectra sources, which remain largely confined within the geometric horizon limits. The broad 21-cm cosmological signal delay spectra in cyan demonstrates that unsmooth spectral signals have a much wider convolving kernal and produce a much wider delay spectra. If analysis is carried out outside of the horizon limits then the foregrounds can be avoided. \textit{Reproduced from Parsons, Pober, Aguirre et al. (2012), ApJ, 756(2):165}.}
    \label{fig:horizon}
\end{figure}

By performing this delay space transform, we are effectively moving into the sidelobes of the 21-cm signal in delay space. The cosmological signal is scattered to high delays whereas the foregrounds are not, allowing the data analysis in that large delay space to be free of foregrounds and foreground removal bias. This method also removes the need for imaging in order to remove the foreground directly, making it suitable for a redundant array with little or no ability to image, but a high sensitivity to the 21-cm power spectrum \cite{Parsons2012ApJ...753...81P}.   

PAPER is a radio interferometer with a highly redundant antenna layout, with multiple baselines of the same length and orientation. Because these multiple baselines all measure the same sky signal, any differences in the signal received would be due to instrumentation, allowing a quick calibration for multiple calibration parameters - `redundant calibration' (e.g. \cite{Ronniy2018AJ....156..285J,Li2018ApJ...863..170L,Dillon2016ApJ...826..181D,Zheng2014MNRAS.445.1084Z,Wieringa1992ExA.....2..203W}). 

PAPER avoided the use of the delay modes dominated by foregrounds and downweighted residual foregrounds using inverse covariance weighting in order to form an upper limit power spectrum measurement \cite{Ali2015ApJ...809...61A}. The latter method of inverse covariance weighting where the covariance is calculated based on the data itself has now been shown to carry the considerable risk of overfitting the EoR data \cite{Cheng2018ApJ...868...26C}. To be clear, despite the retraction of the PAPER-64 results due to power spectrum estimation errors \cite{Ali2018ApJ...863..201A}, the delay space filtering technique remains a promising approach to foreground mitigation.

\subsection{Foreground Subtraction}\label{sec:fgsub}
Foreground subtraction methods all seek to find a model for the observed foregrounds and remove that model from the observed signal, leaving the cosmological signal, instrumental noise and any foreground fitting errors. Foreground removal is usually applied on all scales, meaning that it potentially allows access into the lowest $k_\parallel$ scales where foregrounds traditionally dominate. A caveat of this is that any foreground fitting bias has the potential to affect all scales in the window: foregrounds may remain within the wedge and cosmological signal may be erroneously fitted out within the previously clean EoR window. As an aside, there has been no method so far that can separate out the cosmological 21-cm signal entirely by itself, separate from instrumental noise. Currently when the foregrounds are subtracted or avoided the noise and cosmological signal are still mixed together in what are often termed the `residuals'. The instrumental noise can be obtained from the data for example by the differencing of very fine bandwidth frequency channels, such that both the foregrounds and EoR signal are smooth. The noise power spectrum can then be removed from the residual power spectrum to form the recovered cosmological signal power spectrum. We will now introduce some of the main foreground subtraction techniques.

\subsubsection{Polynomial Fitting and Global Experiments}
\label{sec:poly}

\begin{figure}
\begin{center}
    \includegraphics[width=0.4\textwidth]{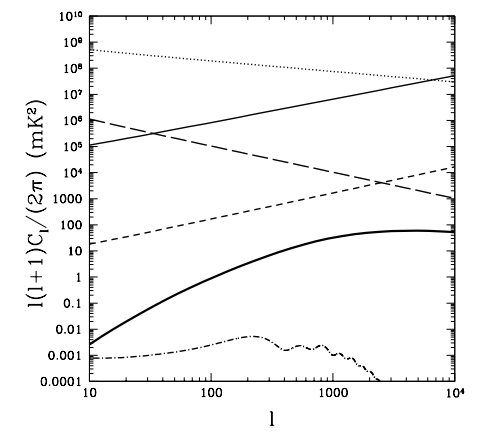}
    \includegraphics[width=0.4\textwidth]{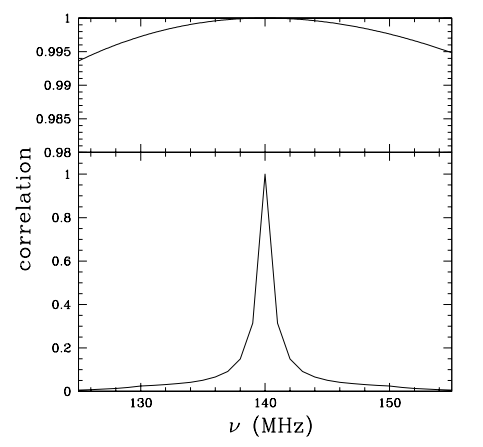}
\end{center}
    \caption{Left: The 2D power spectrum at 140MHz for the cosmological signal (thick, solid), point sources (thin, solid), Galactic synchrotron (thin, dotted), extra-Galactic free-free (thin, dash), Galactic free-free (thin, long dash) and the CMB (dot-dash). The cosmological signal is dominated by foregrounds at all scales, such that separation based purely on spatial differences in not feasible. Figure taken from Fig. 5 of \cite{Santos2005ApJ...625..575S}. Right: The simulated frequency correlations for the foregrounds (top) and cosmological signal (bottom). This plot shows how the correlation between frequency slices (with the comparison made to a slice at 140 MHz), drops off with increasing frequency separation. The foregrounds are highly frequency coherent, whereas the cosmological signal is significantly less so. \textit{Reproduced from Santos, Cooray and Knox (2005), ApJ, 625(2):575--587}.}
    \label{fig:santos_spatial}
\end{figure}

As we have seen in the first half of this chapter, the astrophysical foregrounds are 3-5 magnitudes brighter than the cosmological 21-cm signal and so, by magnitude alone, appear to be the most ominous obstacle to the first detection. Despite, or perhaps because of, their overwhelming magnitude they are well constrained, following power laws with known indices and evolution. The sheer magnitude of the foregrounds means that purely spatial separation, i.e. separation based on only one frequency slice, is not possible: the 21-cm signal and foregrounds are not statistically different enough when only considering spatial scales (see left-hand panel of Fig. \ref{fig:santos_spatial}) \cite{Santos2005ApJ...625..575S,DiMatteo2004MNRAS.355.1053D,Oh2003MNRAS.346..871O,DiMatteo2002ApJ...564..576D}. While separation based purely on spatial scales is not feasible, the high frequency coherence of the foregrounds compared to both the instrumental noise and cosmological signal provides a way to separate out the two signals (foregrounds and both cosmological signal and noise) (see right-hand panel of Fig. \ref{fig:santos_spatial}). 

\cite{Santos2005ApJ...625..575S} and \cite{Zal2004ApJ...608..622Z} exploited the large cross-correlation of the foregrounds in slices at different frequencies to model and remove the foregrounds noting that the frequency coherence was also a useful tool for separation. Polynomial fitting went on to exploit the frequency coherence of the foregrounds across the bandwidth, removing the foregrounds along the line of sight without using any spatial correlation information (e.g. \cite{Bowman2009ApJ...695..183B,Wang2006ApJ...650..529W,Mcquinn2006ApJ...653..815M}). In this method, the foregrounds are modelled by a polynomial function, for example in log-log space such as:

\begin{equation}
\mathrm{\log I} = a_3 +a_2\log\nu + a_1(\log \nu)^2 + ....
\end{equation}

\noindent where $I$ is the brightness temperature of the data, $\nu$ is the frequency of observation and $a_1,a_2,a_3$ are the coefficients which are to be determined in the fit.

Polynomial fitting is a parametric foreground mitigation method. It uses knowledge from simulated foregrounds to tune the coefficients of the polynomial function (e.g. \cite{jelic08}). There are two areas of concern when using this method. Firstly, the effect of the instrument results in a signal which can differ significantly from the frequency-coherent theoretical foreground model (see Section \ref{sec:wedge}). By incorporating weighting according to the amount of information in a particular $uv$ cell, this could possibly be overcome \cite{Liu2009MNRAS.398..401L,Bowman2009ApJ...695..183B}. The second area of concern was that the success of the method relies heavily on having an accurate model for the foreground signal. There are many more instrumental effects than the frequency dependence of the beams, for example polarization leakage (e.g. \cite{Nunhokee2017ApJ...848...47N,Asad2015MNRAS.451.3709A}) and excess instrumental noise \cite{Patil2016MNRAS.463.4317P}. \cite{Wang2013ApJ...763...90W} demonstrated that polynomial removal across the EoR frequency band resulted in significant signal loss when using simulations of complex foregrounds, though they also showed that by fitting a polynomial simultaneously in smaller bandwidth segments this signal loss could be mitigated. Polynomial removal is now rarely used within the interferometric experiments with the exception of the upper limit from GMRT \cite{Paciga2011MNRAS.413.1174P} which used a similar philosophy to remove their foregrounds, albeit by applying a piecewise linear function, as opposed to a polynomial function. 

Aside from interferometric experiments, polynomial fitting does have a prominent place in global EoR experiments (e.g. \cite{Singh2018ApJ...858...54S,Bowman2018Natur.555...67B}) which, due to the coherence of the 21-cm global signal over frequency, means that so far all the more sophisticated methods of foreground mitigation have been unworkable on global simulation and data. For example, the very small number of lines of sight observed by a single global experiment mean that there is not enough spatial information for some non-parametric methods to work.

The Experiment to Detect the Global EoR Signature (EDGES) detection \cite{Bowman2018Natur.555...67B} used a five term polynomial based on the properties of the foregrounds and ionosphere, incorporating the actions of the instrument into their foreground model. The level of accuracy of this method has since questioned however, with the results showing dependence on the description of the foregrounds \cite{Bradley2019ApJ...874..153B,Hills2018Natur.564E..32H}. Overall, polynomial fitting correctly exploits the foreground coherence but it is vulnerable to unknown systematics and unexpected foreground signals. For global experiments there is currently no other option, but for interferometric experiments the methods in the following section provide an alternative.

\subsubsection{Non-parametric foreground removal}
\label{sec:nonpar}
The concern that the instrument might introduce complex spectral structure into the foreground signal has driven research into foreground mitigation methods which rely less on a strongly constrained foreground model. Wp smoothing \cite{Harker2009MNRAS.397.1138H} fits a function along the line of sight whilst penalising the ``Wendepunkt", inflection points, that give the method its name. Unlike polynomial fitting, the function is permitted to be rough but inherently favours the more smooth models. Wp smoothing is applied along each line of sight individually and so spatial correlations of the foregrounds are not utilised in making the foreground fit. The current method employed by the LOFAR EoR pipeline, Gaussian Process Regression (GPR) \cite{Mertens2018MNRAS.478.3640M} also relies purely on spectral information. GPR models the foregrounds, mode mixing components, 21-cm cosmological signal and noise by Gaussian Processes, allowing a clear separation and uncertainty estimation (see Fig. \ref{fig:mertens_comp}). GPR does not require specification of a functional form for each component but instead allows the data to find its own model, while taking into account the covariance structure priors incorporated by the user. This allows a certain level of control, for example splitting the foreground covariance into a smooth intrinsic foreground model and an unsmooth mode mixing component, while still not imposing a strict level of smoothness or a parametric form on the data. 

\begin{figure}
\begin{center}
    \includegraphics[width=0.9\textwidth]{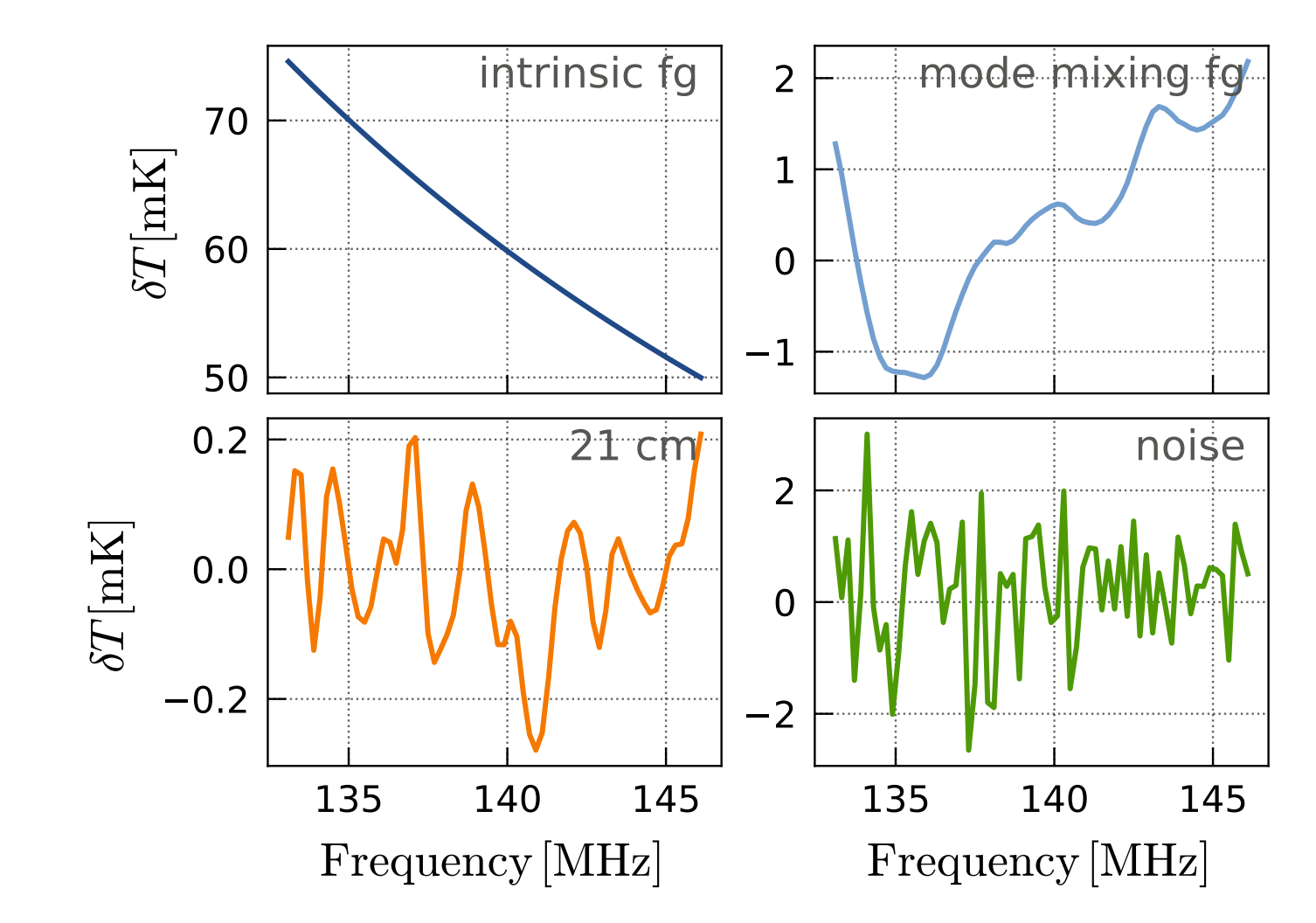}
\end{center}
    \caption{Simulated components of the observed signal, demonstrating that the smooth foreground signal is accompanied by an unsmooth mode mixing signal. GPR models each of these foreground components separately, making use of prior information about each component in the form of covariance functions. \textit{Reproduced from Mertens, Ghosh and Koopmans, (2018), MNRAS, 478(3):3640--3652}.}
    \label{fig:mertens_comp}
\end{figure}

 Blind Source Separation (BSS) methods have been used in Cosmic Microwave Background experiments \cite{PlanckI2018arXiv180706205P,PlanckIV2018} and their application to EoR data is a natural evolution. BSS methods are used across a wide range of fields in order to separate mixed signals into independent components. The data can be expressed in terms of the mixing model:

\begin{equation}
\mathbf{X} = \mathbf{A}\mathbf{S} + \mathbf{N}
\end{equation}

\noindent where $\mathbf{X}$ is the observed signal, $\mathbf{S}$ are the independent components of that signal, $\mathbf{N}$ is the noise and $\mathbf{A}$ is a matrix determining how the components are mixed, the `mixing matrix'. For an observation of $m$ frequency channels each constituting $t$ pixels and a foreground model of $n$ independent foreground components, the dimensions of these quantities are $\mathbf{X}$[$m$,$t$], $\mathbf{S}$[$n$,$t$], $\mathbf{N}$[$m$,$t$] and $\mathbf{A}$[$m$,$n$].

When this framework is applied to EoR data, the foregrounds are contained within $\mathbf{S}$[$n$,$t$] while the cosmological signal is contained along with the instrumental noise in $\mathbf{N}$[$m$,$t$]. The independent components of the foreground model are not directly related to the Galactic synchrotron, Galactic free-free and extragalactic foregrounds, but instead each independent component is potentially a mixture of all these physical foregrounds. This leaves the user without a physically motivated choice for the number of independent components, so that the number must be chosen empirically based on simulated data. Once a foreground model $\mathbf{A}\mathbf{S}$ has been determined this can then be subtracted from the observed signal, leaving the residual data as with the other methods.

\begin{figure}
\begin{center}
    \includegraphics[width=0.9\textwidth]{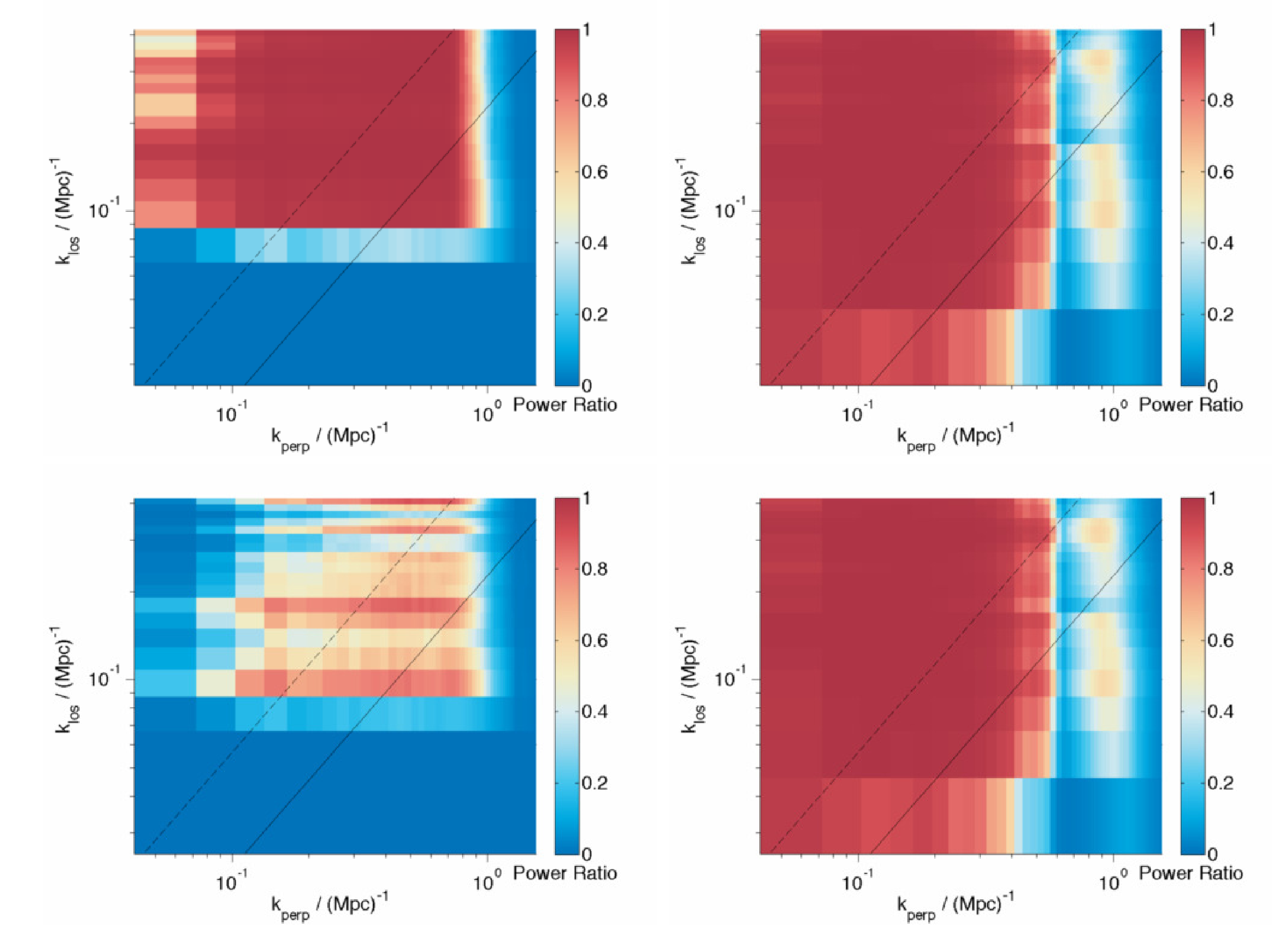}
\end{center}
    \caption{The left column shows the ratio of the simulated components, (cosmological signal / (cosmological signal + foregrounds)), demonstrating that the area of the window free from foreground contamination is small when the foregrounds are unsmooth. The top row is where the foreground model has a random wiggle along the line of sight equal in magnitude to 0.1$\%$ of the foreground signal. The bottom row shows a 1$\%$ wiggle. On the right is the same ratio but with foreground fitting errors after foreground removal by GMCA instead of the simulated foregrounds, demonstrating that the method can open up the EoR window significantly even when the smoothness of the foregrounds is under threat. \textit{Reproduced from Chapman, Zaroubi et al. (2016), MNRAS, 458(3):2928--2939}.}
    \label{fig:Chap_window}
\end{figure}

The two BSS methods introduced for use on EoR data differ by their definition of independence. FastICA \cite{Chapman2012MNRAS.423.2518C,hyvarinen2004independent,hyvarinen1999fast} is a long-established independent component analysis technique which uses statistical independence to separate out the foreground components. FastICA constrains the different components by maximizing the negentropy of the signal components, utilising central limit theorem which states that the more independent components a signal contains, the more Gaussian the probability distribution function of that signal will be. In contrast, GMCA \cite{Bobin2016AA...591A..50B,bobin2015sparsity,Chapman2013MNRAS.429..165C,Bobin2008StMet...5..307B} is a method developed for use on CMB data that uses morphological diversity to separate out components. GMCA assumes that the data is represented in a sparse manner which can be achieved by a wavelet decomposition. With the independent components unlikely to have the same few non-zero basis coefficients in wavelet space, the method is able to separate out the components according to the differing sparse basis coefficient values. As with FastICA, we actually care little for the independent components individually, it is the combination of those as a whole which form the foreground model, with the method naturally separating out the decoherent noise and cosmological signal. In simulation both these methods have behaved well, opening up the EoR window into the lowest scales even when subjected to unsmooth foreground simulations, Fig. \ref{fig:Chap_window}. GMCA was used to achieve the current LOFAR upper-limit \cite{Patil2017ApJ...838...65P} but since then has not been able to remove the foregrounds down to the same level as, for example, GPR \cite{Mertens2018MNRAS.478.3640M}. The reason for this remains unknown and a full comparative analysis is currently underway. \cite{Mertens2018MNRAS.478.3640M} also expressed concern that because BSS methods are not based on defining the components in a statistical framework relating to the contributions from foregrounds and mode-mixing, they are not easily assessed for uncertainty and physical meaning. The blind methods are very useful as a separate check on results from what are extremely complex experiments, with many unknown unknowns. There is scope to move these methods towards a more parametric framework, perhaps constraining the mixing matrix columns according to the first-hand knowledge about the instrumental effects and foregrounds we have built up from the pathfinder telescopes. This is a similar philosophy as introduced by \cite{Bonaldi2015MNRAS.447.1973B} in Correlated Component Analysis (CCA). While still based on a mixing matrix framework, CCA is a parametric method which constrains the mixing matrix to represent power law behaviour over frequency, fixing the spectral index for a Galactic free-free contribution explicitly. 

While Wp smoothing, GMCA, GPR and FastICA are all labelled non-parametric in the literature, it is important to note than none of them are fully blind or indeed fully non-parametric. Each of them require the selection of parameters to define the fit: whether it is the smoothing parameter in Wp smoothing, or the number of independent components in GMCA and FastICA. So far these parameters have been chosen based on minimizing the foreground fitting error on simulated data, where the foreground model is known. A more robust method is to implement a Bayesian model selection model, as GPR does already. In addition, \cite{Gleser2008MNRAS.391..383G} developed a method based on the Bayesian maximum a posteriori probability (MAP) formalism, assuming priors for the smoothness of the contaminating radiation and for the correlation properties of the cosmological signal and \cite{Zhang2016ApJS..222....3Z} introduced HIEMICA (HI Expectation Maximization Independent Component Analysis), an extension of ICA with a fully Bayesian inference of the foreground power spectra, allowing their separation from the cosmological signal power spectra. Machine learning has also been applied in an effort to seek a foreground model defined by the data itself \cite{Li2019MNRAS.485.2628L}. There are now a multitude of non-parametric foreground subtraction methods available which have each proved their own principle on simulated, and in the case of GPR and GMCA, observed data. Now we know the constraints of the instrument much better, work on the relative advantages and disadvantages of all these approaches are a logical next step.

\subsection{Residual Error Subtraction}
The final stage of foreground mitigation is residual error subtraction \cite{Morales2006ApJ...648..767M,Morales2004ApJ...615....7M}. The residual foreground mitigation errors from the previous two stages (bright source subtraction and diffuse foreground mitigation) produce distinct shapes in the spherical power spectrum, Fig. \ref{fig:ressub}. One can take the spherical power spectrum of the residual data and apply a multi-parameter fit according to the foreground residual and EoR template power spectrum. This allows a final cleaning of residual foreground contamination. \cite{Morales2006ApJ...648..767M} also notes that ``because the residual error subtraction relies on the statistical characteristics of the subtraction errors, the foreground removal steps become tightly linked and we must move from focusing on individual subtraction algorithms to the context of a complete foreground removal framework." This statement leads us neatly to the conclusion of this chapter.

\begin{figure}
\begin{center}
    \includegraphics[width=0.9\textwidth]{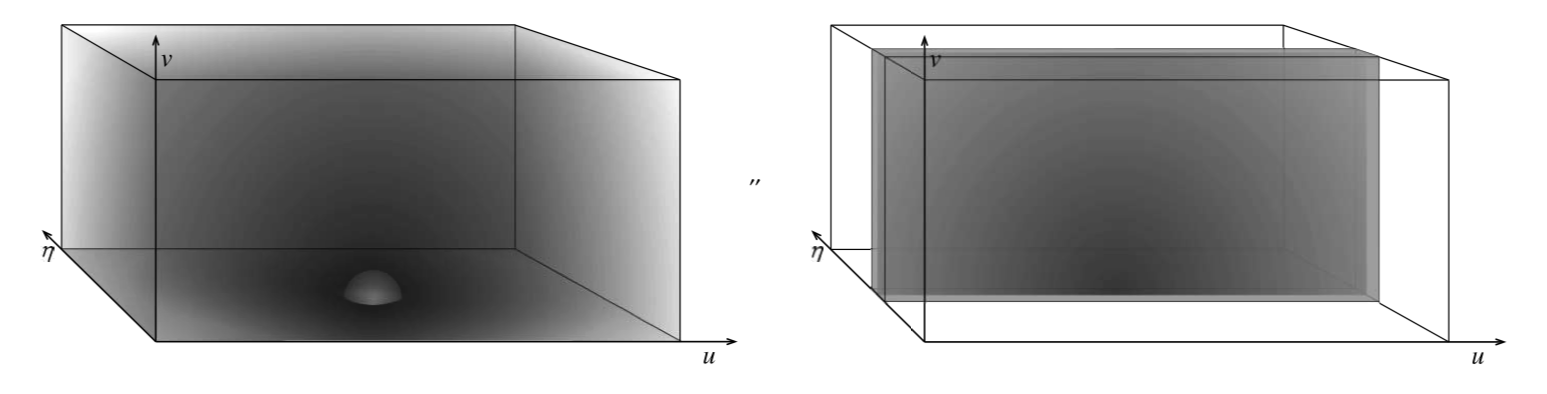}
\end{center}
    \caption{The 3D spherical power spectrum of the EoR signal (left), and an example residual foreground signal template (right), where zero is at the centre of the bottom face of the cuboid. The foreground signal displays a separable-axial symmetry while the EoR signal has a symmetric power spectrum. This contrast allows a further separation stage in order to clean the foreground fitting errors which have accumulated from the previous two stages of bright source subtraction and diffuse foreground mitigation. \textit{Reproduced from Morales, Bowman, and Hewitt (2006), ApJ, 648(2):767--773}.}
    \label{fig:ressub}
\end{figure}

\subsection{Polarization leakage}\label{sec:leakage}
One of the challenges in calibration is to minimise leakage of polarization signals in total intensity. Otherwise, the polarization leakage can contaminate the cosmological 21-cm signal. 
A level of contamination depends strongly on characteristics of a radio telescope, its calibration strategy, and of polarized emission itself.

Antennas in the low-frequency radio telescopes are dipoles. Dipoles usually come in pairs. In each pair dipoles are orthogonal to each other and each dipole is sensitive to a certain polarization. Since antennas are also fixed to the ground, it is not possible to preform observations like with the traditional dish-like radio telescopes, where the tracking  is done by steering the dish. Here, the sources are tracked by the beam-forming or simply the observation is done in a drift-scan mode. Depending on the position of the sources in the sky, the sources will see different projections of dipoles. If this geometrical projection  is not corrected during the calibration, or the modelling of and correction for the beam polarization is not accurate, polarized signals can leak to total intensity and vice versa. 

\begin{figure}[!t]
   \centering	
   \includegraphics[width=0.8\textwidth]{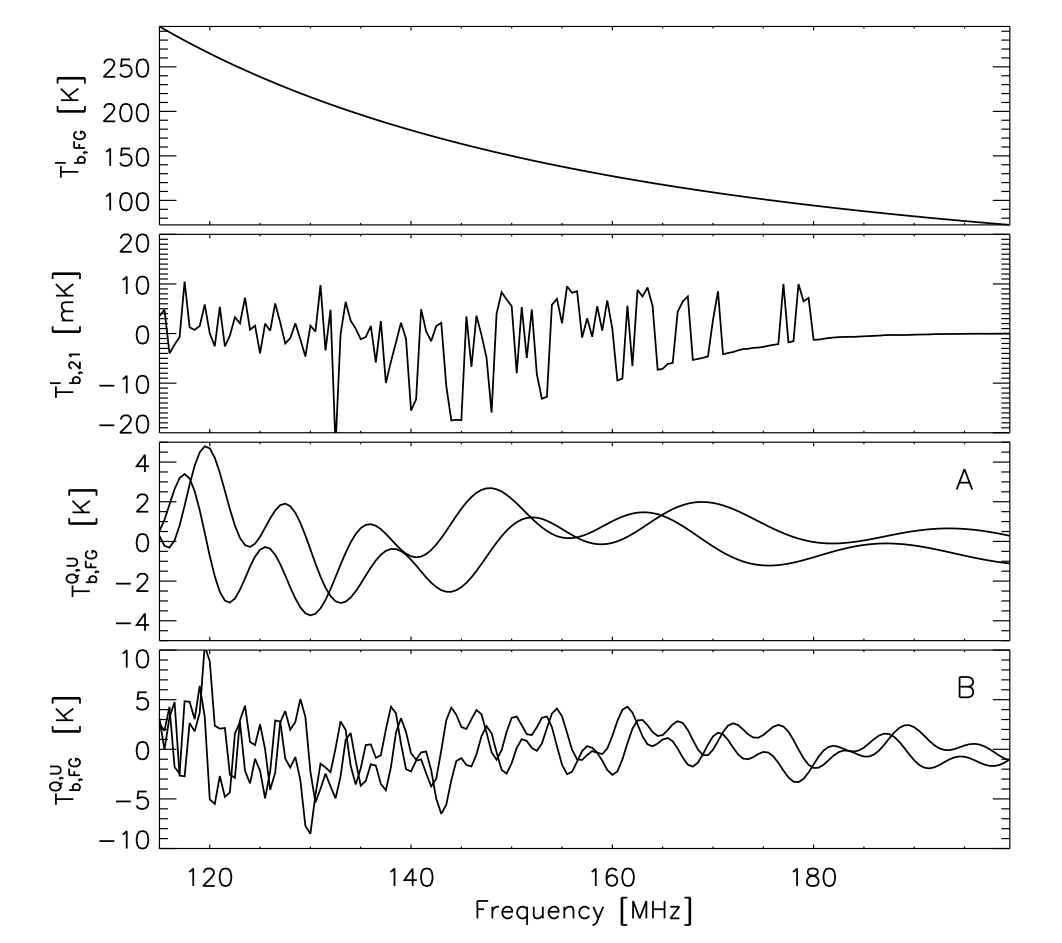}
    \caption{Galactic synchrotron emission given as a function of frequency in total intensity (Stokes I, $T^I_{b,FG}$) and polarization (Stokes Q and U, $T^{Q,U}_{b,FG}$). The spectra are generated by Jeli\'c simulations \cite{jelic10, jelic08}. Polarized emission can have a very complex frequency dependence compared to the plain power-law behaviour in the total intensity, due to distinct Faraday rotation and depolarization at low-radio frequencies (see Sec.~\ref{sec:polarfg}). If polarized emission consists of the multiple Faraday components and/or if some of the components are at $\Phi\gtrsim15~{\rm rad~m^{-2}}$ (example B vs. A) this can create a leaked signal in total intensity that looks like the cosmological 21cm signal ($T^I_{b,21}$, in this case generated by 21cmFAST \cite{mesinger11}).  }
\label{fig:leakage}
\end{figure}

Since the polarized emission from the Milky Way can have a very complex frequency dependence, a leakage of this signal to the total intensity can contaminate the cosmological 21-cm signal, making extraction and analysis more demanding (\cite{jelic10, moore13, spinelli18} and see Fig.~\ref{fig:leakage}). A number of studies addressed this problem for different low-frequency radio telescopes: LOFAR \cite{asad18, asad16, asad15},  MWA \cite{sutinjo15} and PAPER \cite{kohn16, nunhokee17}.  Although the assessed polarization leakage in these studies is not limiting current  observations, it will become relevant once we reach a better sensitivity in the data. This will be especially the case for future 21 cm experiments, like HERA and SKA. 

Most of the observed structures appear at Faraday depths $|\Phi|\lesssim15~{\rm rad~m^{-2}}$, which measures the amount of Faraday rotation by
intervening interstellar medium (see Sec.~\ref{sec:polarfg}). Relatively small Faraday depths indicate polarized emission that fluctuates along frequency on scales  larger than 
the expected cosmological 21-cm signal  in total intensity (e.g. \cite{moore13}). Thus, associated leaked signals can be in principle mitigated, as it was shown in the case of a simple and  thin Faraday screen \cite{geil11}. On the contrary, polarized emission at Faraday depths $|\Phi|\gtrsim15~{\rm rad~m^{-2}}$ can introduce frequency dependent signals, which if leaked can resemble the cosmological signal and make foreground mitigation difficult.  Prior to the CD/EoR observations, it is therefore important to asses the properties of Galactic polarized emission in targeted region of the sky.

\section{Conclusions}
The foreground emission of our Galaxy and extragalactic radio sources dominates over the whole frequency range of the cosmological 21cm experiments. In order to mitigate the foreground emission from the data we need to study and constrain its properties in great details. Thanks to the observations with LOFAR and MWA this is becoming possible.

The current EoR experiments are now modelled and constrained to an excellent degree but during that process there has been a blurring of boundaries between the analysis modules. The calibration stage, once assumed to mitigate foreground point sources only, can erroneously suppress diffuse foregrounds \cite{Patil2016MNRAS.463.4317P} and the mode-mixing of the instrument has required more complex modelling as wide-field effects have become apparent \cite{Thyagarajan2015ApJ...807L..28T}. There are a promising number of foreground mitigation techniques now available providing the necessary diversity of pipelines necessary for verifying the first detection. So far, there has not been a wide-reaching comparison of all of these methods or a complete assessment of their strengths and weaknesses for recovery of the different aspects of the EoR signal such as power spectra or images. Foreground subtraction, suppression and avoidance are now used in combination in the experimental pipelines and the further development of the best combination for these methods will provide an exciting area of research in the next decade.

\bibliographystyle{plain}
\bibliography{References}